# Electronic Structures and Structural Evolution of Hydrogenated Graphene Probed by Raman Spectroscopy

Zhiqiang Luo[†‡], Ting Yu[†], Zhenhua Ni[§], Sanhua Lim[‡], Hailong Hu[†], Jingzhi Shang[†], Lei Liu[†], Zexiang Shen[†*], and Jianyi Lin[‡*]

Division of Physics and Applied Physics, School of Physical and Mathematical Sciences, Nanyang Technological University, Singapore 637371

Applied Catalysis, Institute of Chemical and Engineering Sciences, Singapore 627833

Department of Physics, Southeast University, Nanjing, China 211189

* Corresponding authors email: Zexiang@ntu.edu.sg; lin_jianyi@ices.a-star.edu.sg

† Nanyang Technological University
‡ Institute of Chemical and Engineering Sciences
§ Southeast University

**ABSTRACT:**

The electronic structures and structural evolution of hydrogenated graphene are investigated by Raman spectroscopy with multiple excitations. The excitation energy dependent saturation effect on the ratio of integrated intensities of D and G modes ($I_D/I_G$) is revealed and interpreted by a D band active model with D band Raman relaxation length and photo-excited electron/hole wavelength as critical length scales. At low hydrogen coverage, the chemisorbed H atoms behave like defects in *sp2* C=C matrix; while for a high hydrogen coverage, the *sp3* C-H bonds become coalescent clusters, resulting in confinement effect on the *sp2* C domains. Electronic structure changes caused by varying hydrogen coverage are evidenced by excitation energy dependent red shift of D and 2D bands. Our results provide a useful guide for developing applications of hydrogenated graphene, as well as using Raman spectroscopy for quick characterization in further exploring other kinds of graphene derivatives.

## 1. Introduction

Graphene, a single atomic layer of graphite, has attracted great attentions because of its exceptional physical properties and its potential applications.[1] However, graphene as a semi-metal has been limited from some device applications. Opening a bandgap or modifying the intrinsic semi-metallic property of graphene is a crucial step to the wider applications of graphene in electronics and photonics.[1,2] Recently, grafting of various atoms or functional groups on graphene, namely graphene derivatives fabricated by chemical surface modification, has been attracted tremendous attention to exploit their possibilities and efficiency in tuning the electronic band structures of graphene.[3-7] For example, exposing graphene to atomic hydrogen was shown to generate C-H bonds on its surface, resulting in an increase of the *sp³* hybridization and a dramatic alteration of their local electronic structure, which rendered a transition from metal-like conductor to nearly ideal two dimensional insulator.[4] Through patterned adsorption of atomic hydrogen onto the Moiré superlattice positions of graphene grown on an Ir (111) substrate, Balog *et al*. demonstrated the existence of a confinement-induced bandgap opening in the hydrogenated graphene by angle-resolved photoemission spectroscopy (ARPES) investigation.[6] The electronic structure of hydrogenated graphene should strongly depend on the hydrogen coverage.[6] Based on theoretical prediction, a fully (double sides) hydrogenated graphene, i.e. graphane, is a semiconductor with a large band gap.[8] For single side hydrogenation, a band gap insulating behavior has been predicted at high coverage,[9] while at lower coverage, a localized insulating behavior has been observed at low temperature.[4] Moreover, at ultra low hydrogen coverage, a metal-insulator transition was observed on hydrogenated epitaxial graphene on SiC substrate.[10]

Raman spectroscopy has been widely applied to exploit the structural and electronic properties of graphene, including layer numbers, stacking order, strain effect, and doping concentration.[11-14] In this paper, we present systematic Raman spectroscopic investigations on hydrogenated graphene with different amount of hydrogen coverage, which are controlled by modulating hydrogen plasma treatment dose with varied parameters such as plasma power, $H_2$ pressure and process duration.[5] The D, G and 2D bands of hydrogenated graphene show significant dependence on the hydrogen coverage as well as the excitation energies, providing valuable information about the evolution of structural and electronic properties of graphene with increasing hydrogen coverage, which would be a useful guide for the application of hydrogenated graphene.

## 2. Experimental Methods

Graphene on Si wafer substrate with 285 nm $SiO_2$ cap layer were prepared by mechanical cleavage from highly ordered pyrolytic graphite (HOPG). Optical microscope was used to locate the graphene samples. The graphene samples were directly immersed in hydrogen plasma at 10W, 1 Torr (133 Pa) with different duration. The hydrogen plasma was ignited between two metallic parallel-plate electrodes of 20 cm in diameter and 4 cm separation in a capacitively coupled radio frequency (13.56 MHz) PECVD reactor.[5] Raman spectra were recorded with different excitation lasers: 2.71 eV (457 nm), 2.54 eV (488 nm), and 2.33 eV (532 nm) with WITEC CRM200 Raman system; 1.96 eV (633 nm) and 1.58 eV (785 nm) with Renishaw inVia Raman system. The laser power on sample is kept below 1 mW to avoid possible laser-induced heating.

## 3. Results and Discussion

Raman spectra of pristine, hydrogenated, and dehyrogenated graphene under 2.33 eV excitation are shown in Figure 1a. The Raman spectrum of pristine graphene displays two characteristic peaks, G band at ~1580 $cm^{-1}$ and 2D band at ~2670 $cm^{-1}$, which are assigned to the in-plane vibrational mode ($E_{2g}$ phonon at the Brillouin zone center), and the intervalley double resonance scattering of two TO phonons around the $K$-point of the Brillouin zone respectively.[14] After hydrogenation by $H_2$ plasma treatment at 10W, 1Torr, 9 min, three defect induced peaks at 1340, 1620 and 2920 $cm^{-1}$ are observed in Raman spectrum of hydrogenated graphene, which are assigned to D band activated by defects via an intervalley double-resonance Raman process, D' band activated by defects via an intravalley double-resonance process, and the combination mode of D and G modes (D+G band) respectively.[14] The defects in hydrogenated graphene result from the formation of $sp^3$ C-H bonds as well as the breaking of the translational symmetry of $sp2$ C=C network.[4] Commonly the frequency-integrated intensity ratio of D band to G band ($I_D/I_G$) can serve as a convenient measurement of the amount of defects in graphitic materials.[14] The $I_D/I_G$ of the above hydrogenated graphene is as high as 5 (See Fig. 1a), suggesting high defect density induced after hydrogenation.[4,5] After vacuum annealing of above hydrogenated graphene at 500°C for 30 min, almost all of the defect-related Raman bands (D, D', D+G) can be eliminated, as shown in Raman spectrum of dehydrogenated graphene, indicating the defects are mainly $sp^3$ monovalent hydrogen adsorbates, which are reversible and can be thermally healed to restore the original graphene lattice.[4]

According to double-resonance process, the excitation energy dependence of D and 2D bands can be used to explore the electronic structures of graphene after hydrogenation.[14] Figure 1b shows

Raman spectra of the hydrogenated graphene, which was treated with 10W, 1Torr, 3 min $H_2$ plasma, using three different excitations. The spectra are normalized respect to the G band intensity. Note that both the peak positions and the frequency-integrated intensity ratios ($I_D/I_G$ and $I_{2D}/I_G$) of D and 2D bands are strongly dependent on the excitation laser energy ($E_L$). The $I_D/I_G$ is 1.6, 3.1 and 6.3 for 2.71, 2.33 and 1.96 eV excitation, respectively, which follows an $E_L^{-4}$ relation. The inverse proportion of $I_D/I_G$ to the fourth power of the laser energy was previously reported in a Raman study of nanographite.[15,16] Based on the calculation of Raman scattering theory, matrix elements associated with the double resonance Raman processes of D band show an $E_L$ dependence of $E_L^{-4}$ for nanographite.[16] For 2D band, it is predicted to have an excitation energy dependence of $E_L^{-3}$.[17] However, the intensity ratio of $I_{2D}/I_G$ of pristine graphene keeps nearly unchanged with increasing excitation energy and increase slightly when excited with high excitation energy (2.41 and 2.71 eV).[17] Similarly, the $I_{2D}/I_G$ of hydrogenated graphene are 3.6, 2.8 and 3 for 2.71, 2.33 and 1.96 eV excitation respectively, with a slight increase when excited with higher excitation energy (2.71 eV). The deviation from the $E_L^{-3}$ relation in pristine and hydrogenated graphene may caused by other electron scattering processes in addition to the electron-phonon scattering in double-resonance process of 2D band, which will be discussed in detail latter.

Following, a series graphene samples with different amounts of hydrogen coverage were prepared and studied with different Raman excitation energies. The $I_D/I_G$ ratio and $I_D/I_G$ times $E_L^4$ are plotted as a function of hydrogen coverage (or hydrogen plasma treatment dose, defined by W*Torr*min) in Figures 2a and 2b, respectively. Under the excitation of 2.71 eV, 2.54 eV and 2.33 eV, the ratio $I_D/I_G$ is firstly proportional to both the hydrogen coverage and the inverse fourth power of the laser energy ($E_L^{-4}$), but becomes inversely proportional to the hydrogen plasma treatment dose and significantly deviated from the $E_L^{-4}$ relation when the $H_2$ plasma dose is larger than 90W*Torr*min (stage IV as indicated in Figures 2a and 2b). However, in addition to the two increasing and decreasing parts described above, the evolutions of $I_D/I_G$ under 1.96 eV and 1.58 eV excitations show one more part, a saturation stage with no further change for increasing hydrogen coverage. For example, under the excitation of 1.96 eV, the $I_D/I_G$ saturation stage happens when the $H_2$ plasma dose is between 50 and 90 W*Torr*min (stage III as indicated), while the $I_D/I_G$ saturation stage starts at lower $H_2$ plasma dose (plasma dose ≥ 10W*Torr*min, stage II and stage III) when the excitation is 1.58 eV near infrared (NIR) laser. The Raman spectra (recorded using 1.96 eV excitation) of the hydrogenated graphene at each stages were shown in Figure 3 for better understanding of this saturation phenomenon.

The unusual decrease of $I_D/I_G$ with increasing amount of defects (stage IV here) after a $I_D/I_G$ maximum, was also reported by Lucchese *et al.* in Raman spectroscopy investigation of defective graphene induced by $Ar^+$ ion bombardment, and interpreted by a local active model of D band.[18] In their phenomenological model, the D band in Raman spectra was proposed to be mostly contributed by activation area, i.e. the hexagonal lattices proximity to the defect region, while the defect region, whose radius $r_D$ is around 1 nm (revealed by scanning tunneling microscopy study), make less contribution to the D band due to strong structural disorder and breakdown of the hexagonal crystalline structure.[18] The size of the activation area is determined by its radius $r_A = r_D + l$, where $l$ is Raman relaxation length for the resonant Raman scattering of D band.[18-20] The Raman relaxation length $l$ is the average distance traveled by an electron before undergoing inelastic scattering by a phonon, since the phonons of D bands can only become Raman active if the electrons are involved in both the electron-defect elastic scattering and electron-phonon inelastic scattering.[19,20] Upon increasing the defect density, the corresponding activation areas are created independently from others and eventually overlap, and therefore the D band intensity will reach a maximum when the inter-defect distance, i.e. mean distance between defects, $d = r_A + r_D$. If the defect density is large enough, the defect regions start to coalesce and the activation areas with hexagonal lattice shrink, which would significantly reduce the D band intensity.[18] Our case appears to follow this structure evolution trajectory. In stage IV, the inter-defect distances become so small that structural distortion region caused by C-H bonding should coalesce (see Fig. 4). As revealed by scanning tunneling microscopy (STM) investigation of hydrogenated epitaxial graphene grown on SiC substrate, the hydrogenation site in graphene still keeps the hexagonal lattice structure after chemisorption of hydrogen dimmers.[21] The predicted structural distortion region caused by C-H bonding is in the range of a few atoms around the hydrogenation site, much smaller than 1nm.[9] Moreover, different from the fixed defect size induced by Ar sputtering, the size of the hydrogenation site, i.e. $r_D$, keeps enlarging with increasing hydrogenation coverage, since the hydrogen atoms prefer stick together to form hydrogen clusters.[21]

The observed excitation energy dependent $I_D/I_G$ saturation phenomenon can be viewed as a excitation energy dependent $I_D/I_G$ maximum effect, because the $I_D/I_G$ maximum value under 1.58-2.71 eV excitations locate at the hydrogenation dose of 10, 50, 90, 90, 110 W*Torr*min, respectively. The origination of this effect is the fact that D band Raman relaxation length, $l = v_F/(2\gamma)$, shows electron energy dependence.[20] The $v_F$ is Fermi velocity ($v_F = 1.1 \times 10^6$ m/s ≈ 7.3 eV·Å/h in graphene) and the $2\gamma$ is inelastic-scattering rate for a photo-excited electron or hole due to phonon emission.[20,22] According to reference 20, $2\gamma = (\lambda_\Gamma+\lambda_K)E/2$, where $\lambda_\Gamma$ and $\lambda_K$ are dimensionless electron phonon coupling constant, and $E$ is the photo-excited electron or hole energy ($E = E_L/2$). Therefore, $l = 2v_F/[(\lambda_\Gamma+\lambda_K)E]$, which implies larger inter-defect distance and lower defect density for the $I_D/I_G$ maximum at lower photo-excited electron energy, in agreement with our observation. In order to understand the saturation phenomenon after the maximum point, an illustration is shown in Figure 4 for a clear physics picture. When the inter-defect distance $d > 2r_D+l$, $I_D/I_G$ is proportional to defect density (hydrogen coverage) as described in the D band active model proposed by Lucchese et al..[18] When $\lambda_e < d -2r_D < l$, where the $\lambda_e$ is electron wavelength of photo-excited electron or hole and $\lambda_e = v_F/E$, the same photo-excited electron-hole pair can scatter from several defects before emitting a photon, therefore several defects function like a single defect in D band double resonance scattering process, which leads no significant change in D band intensity with different defect density as what happens for stage II and III in our experiment. When $d-2r_D < \lambda_e$, the electron simply cannot distinguish between different defects and defects coalesce together to grow up into one big defect in the electron's "eye", thus such "merger of defects" contributes to the decrease of $I_D/I_G$ in stage IV. For the photo-excited electron with lower energy and larger wavelength, the "merger of defects" process is faster, and therefore results in a rapider drop in $I_D/I_G$, which is consistent with the observation in stage IV. The absence of saturation phenomenon under high excitation energy, for example 2.54-2.71 eV, results from the fact that in graphene with high defect density, D band Raman relaxation length $l$ becomes comparable to the electron wavelength $\lambda_e$. In pristine graphene, the D band Raman relaxation length $l$ for 0.98 eV (1.96 eV/2) electron is around 3 nm, which was deduced in Raman spectroscopy study of graphene edge.[19] In highly defective graphene, Raman relaxation length $l$ should be even smaller due to significantly reduced Fermi velocity around defect points.[23]

The electron energy and inter-defect distance (hydrogen coverage) dependent electron scattering also play important roles in evolution of $I_{2D}/I_G$ ratio. As shown in Figure 2c, a clear tendency shows that the ratio $I_{2D}/I_G$ decreases with increasing hydrogen coverage; meanwhile, there is also a clear transition point between the third and fourth stages. In pristine graphene, the ratio $I_{2D}/I_G$ decreases strongly with increasing doping level due to the additional electron-electron scattering contribution, because the calculated 2D band intensity is proportional to $1/\gamma'^2$, where $2\gamma'$ is the photo-excited electron or hole inelastic scattering rate.[12,24] In hydrogenated graphene, the electron-defect collisions, in addition to the electron-phonon and electron-electron collisions, will contribute to the electron or hole inelastic scattering rate. Therefore, the $I_{2D}/I_G$ decreases with increasing hydrogen coverage, i.e. defect density. When taking account of the dependence of

$I_{2D}/I_G$ on excitation energy, it is found that excitation energy dependence of $I_{2D}/I_G$ are different at different hydrogen coverage. Moreover, contrary to what happened for $I_D/I_G$ in the transition point between stage III and stage V, the drop of $I_{2D}/I_G$ is more drastic when the photo-excited electron (excitation) energy is larger. These observation suggest 2D band Raman scattering is also significantly influenced by energy dependent electron dynamics. The deviation of $I_{2D}/I_G$ dependence on electron energy from the prediction,[17] might be ascribed to both electronic structure dependent electron-electron scattering and inter-defect distance dependent impurity radius in electron-defect scattering.[25,26] Theoretical calculation predicted that resonant scattering of electrons with short-range defects, which are populated in hydrogenated graphene, dramatically increased the scattering cross section and introduced a strong energy dependence.[26]

In stage IV, the peak position of G band up-shifts, as shown in Figure 2d (recorded using 2.33eV excitation), while it keeps nearly unchanged at the first three hydrogenation stages. The stiffening of the G band in the fourth stage might be analogous to the phonon confinement as that found in clustering of the *sp2* phase during the amorphization trajectory of graphite.[27] As there is no obvious up-shift of G band in the first three hydrogenation stages, the photo-excited electron and phonon in the hydrogenated graphene should not be strongly confined. At the first three stages, the hydrogen chemisorped on graphene and form $sp^3$ C-H bonds, working as point defects distributed randomly in the graphene $sp^2$ matrix, while at high hydrogen coverage (stage IV) the *sp3* C-H bonds start to coalesce and form clusters, as described above, and finally $sp^2$ carbon domain were encircled within the $sp^3$ C-H matrix. The coalesced $sp^3$ clusters work similarly to the one-demensional defect structure (grain boundary) in nanocrystalline graphite and nanocrystalline graphene which change and cut the long range phonon interaction.[27,28] Therefore, the electronic structure and phonon dispersion of hydrogenated graphene at the first three stages should be similar to those of graphene, while they change significantly at high hydrogen coverage (stage IV), which is consistent with the observed deviation of the $E_L^{-4}$ dependence of $I_D/I_G$ at stage IV.

These randomly distributed point defects and coalescent defects also have different influence on the full width at half-maximum (FWHM) of D, G and 2D bands, as shown in the inset of Figure 2d (recorded using 2.33eV excitation). Unlike the monotonic increase of FWHM of D, G and 2D band with the decreasing crystallite size in nanographite.[28] the FWHM of D, G and 2D band of hydrogenated graphene evolutes in several stages with increasing hydrogen coverage. At low hydrogen coverage (stage I), the FWHM of G band increases rapidly with increasing hydrogen coverage, resulting from the decrease of phonon lifetime caused by the increasing probability for the phonon-defect scattering. When compared to the FWHM of the first-order Raman G band, the FWHM of D band are broader due to its double-resonance nature of the scattering processes in additional to the phonon lifetime broadening, since a range of phonons with different wave vectors is involved in double-resonance scattering.[29] At medium hydrogen coverage (stage II and III), the FWHM of G and 2D bands increases very slowly with increasing hydrogen coverage and the FWHM of D band keeps nearly unchanged; while the corresponding $I_D/I_G$ in this region increase with increasing hydrogen coverage. The deviation from the increasing tendency of FWHM with increasing $I_D/I_G$, which was demonstrated in Raman spectroscopy study of nanographite, may originate from the different kinds of defects.[28] At high hydrogen coverage stage (stage IV), the FWHM of D, G and 2D bands increases rapidly with increasing hydrogen coverage, because the coalescent defects cut the long range phonon interaction.

Figures 5a and 5b display the excitation energy dependence of D and 2D peak positions, respectively, where the D and 2D bands show linear energy dispersion with rates of around 50 cm$^{-1}$/eV and 100 cm$^{-1}$/eV, which are similar to those in graphite.[14] Note that, the energy dispersion rate of D and 2D band peak position for the hydrogenated graphene decrease with increasing hydrogen coverage. For instance, the 2D band energy dispersion rate of graphene treated with 10W*Torr*min hydrogen plasma dose is around 102 cm$^{-1}$/eV, while the energy dispersion rate of the graphene treated with 90W*Torr*min plasma dose decreased to 92 cm$^{-1}$/eV. It is also interesting to notice that, with a significant red-shift of D and 2D band peak position for all the excitation energies, the energy dispersion rates of D and 2D band become back to ~50 cm$^{-1}$/eV and ~100 cm$^{-1}$/eV at high hydrogen coverage. At low hydrogen coverage, both the electronic band structure and phonon dispersion of hydrogenated graphene are supposed to be similar to those of pristine graphene. However the red-shift of D and 2D band and the change in 2D band peak energy dispersion rate with increasing hydrogen coverage imply that both the electronic band structures and the phonon dispersion should be modified at medium hydrogen coverage (stage II and III), since the dispersion of D and 2D bands with laser energy is proportional to $v_{Ph}/v_F$, where $v_{Ph}$ is the slope of the transversal optical phonon branch going through K and $v_F$ is the Fermi velocity, i.e. slope of the π band energy dispersion near *K*-point.[30] Since a recent STM/STS measurement of defective graphene demonstrated the significant reduction in the Fermi velocity, the $v_{Ph}$ should also decrease with increasing hydrogen coverage.[23] At high hydrogen coverage, there should be a band gap at low energy level near the *K*-point as predicted by theoretical calculation,[9,31] therefore the up shift of the π electron band would cause the significant red-shift of D and 2D band for all the excitation energies.[31] The detailed interpretation of the influence of hydrogen coverage on the energy dispersion of D and 2D band need further intensive theoretical investigation and electronic structure characterization by other experimental methods.[6,32]

## 4. Conclusion

In summary, the electronic structures and structural evolutions of hydrogenated graphene with different amounts of hydrogen coverage are investigated by Raman spectroscopy with multiple excitation. At low hydrogen coverage the chemisorbed H atoms behave like defects in $sp^2$ C=C matrix; while at high hydrogen coverage, the $sp^3$ C-H bonds become coalescent clusters, resulting in confinement effect on the $sp^2$ C domains. The energy dispersion of D and 2D band show clear dependence on hydrogen coverage, indicating the change in electronic structure of hydrogenated graphene. Moreover, a D band active model with both the D band Raman relaxation length and the photo-excited electron/hole wavelength as critical length scales was developed for interpretation of the excitation energy dependent $I_D/I_G$ saturation effect. Our systematic Raman spectroscopic investigations of the hydrogenated graphene provide a useful guide for developing applications of hydrogenated graphene, as well as using Raman spectroscopy for quick characterization in further exploring other kinds of graphene derivatives.

## Acknowledgements

Yu Ting acknowledges the support by Singapore National Research Foundation under NRF RF Award No. NRFRF2010-07 and MOE Tier 2 MOE2009-T2-1-037.

**Figure captions:**

Figure 1. (a) Raman spectra of pristine graphene, hydrogenated graphene (10 W, 1 Torr, 9 min; i.e. 90 W*Torr*min) and dehydrogenated graphene (vacuum annealing at 500°C for 30 min) excited by 2.33 eV laser. (b) Raman spectra of hydrogenated graphene (10 W, 1 Torr, 3 min; i.e. 30 W*Torr*min) excited by 2.71 eV, 2.33 eV and 1.96 eV lasers.

Figure 2. (a) The evolution of D and G band intensity ratio ($I_D/I_G$) with increasing hydrogenation dose in Raman spectra excited by five lasers; (b) The evolution of $(I_D/I_G)*E_L^4$ with increasing hydrogenation dose in Raman spectra excited by five lasers; (c) The evolution of 2D and G band intensity ratio ($I_{2D}/I_G$) with increasing hydrogenation dose in the Raman spectra excited by four visible lasers; (d) The evolution of G band peak position with increasing hydrogenation dose in Raman spectra excited by 2.33 eV laser; the insert shows the FWHM evolution of D, G and 2D bands with increasing hydrogenation dose in Raman spectra excited by 2.33 eV laser.

Figure 3. Raman spectra of hydrogenated graphene with different hydrogen coverage. The excitation energy is 1.96 eV.

Figure 4. Illustration of D band active model with both the D band Raman relaxation length and the photo-excited electron/hole wavelength as critical length scales.

Figure 5. (a) The D band peak position of graphene with different hydrogen coverage as a function of excitation energy; (b) The 2D band peak position of graphene with different hydrogen coverage as a function of excitation energy.

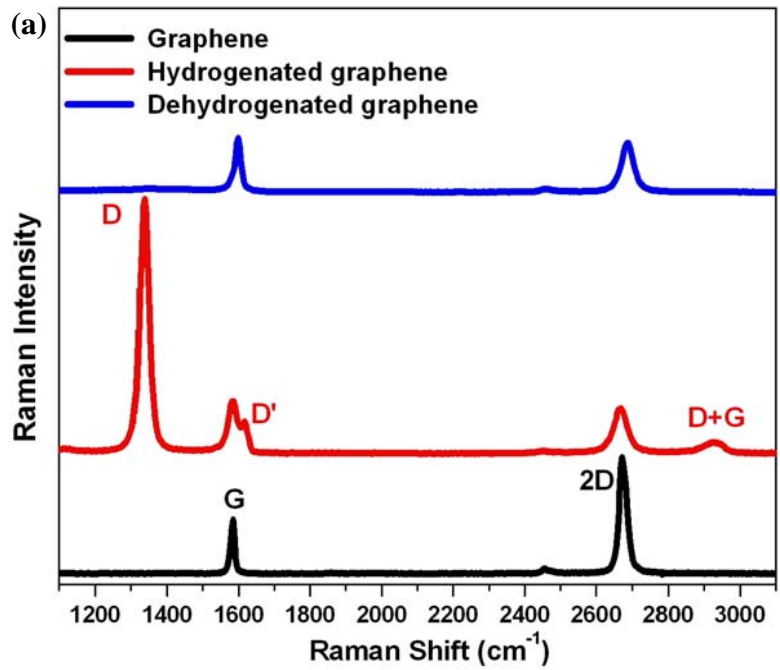

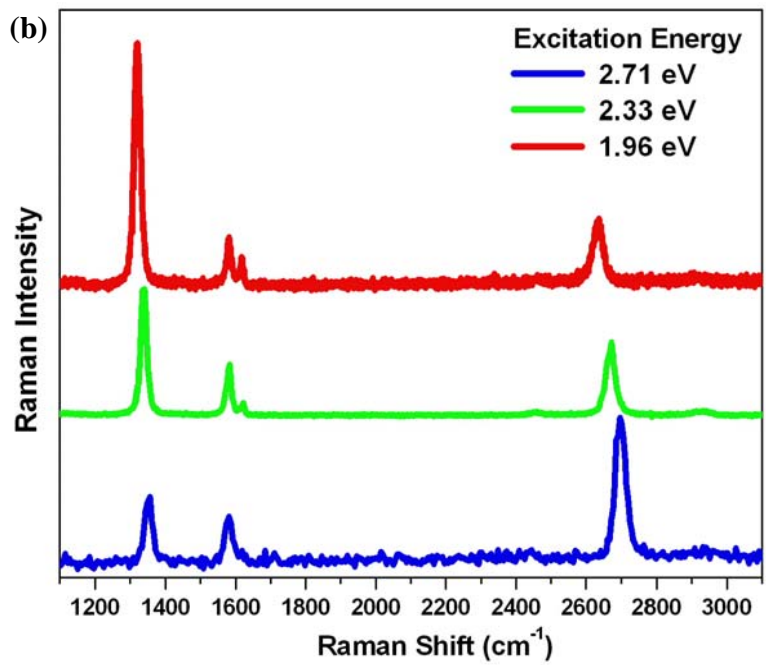

**Figure 1.**

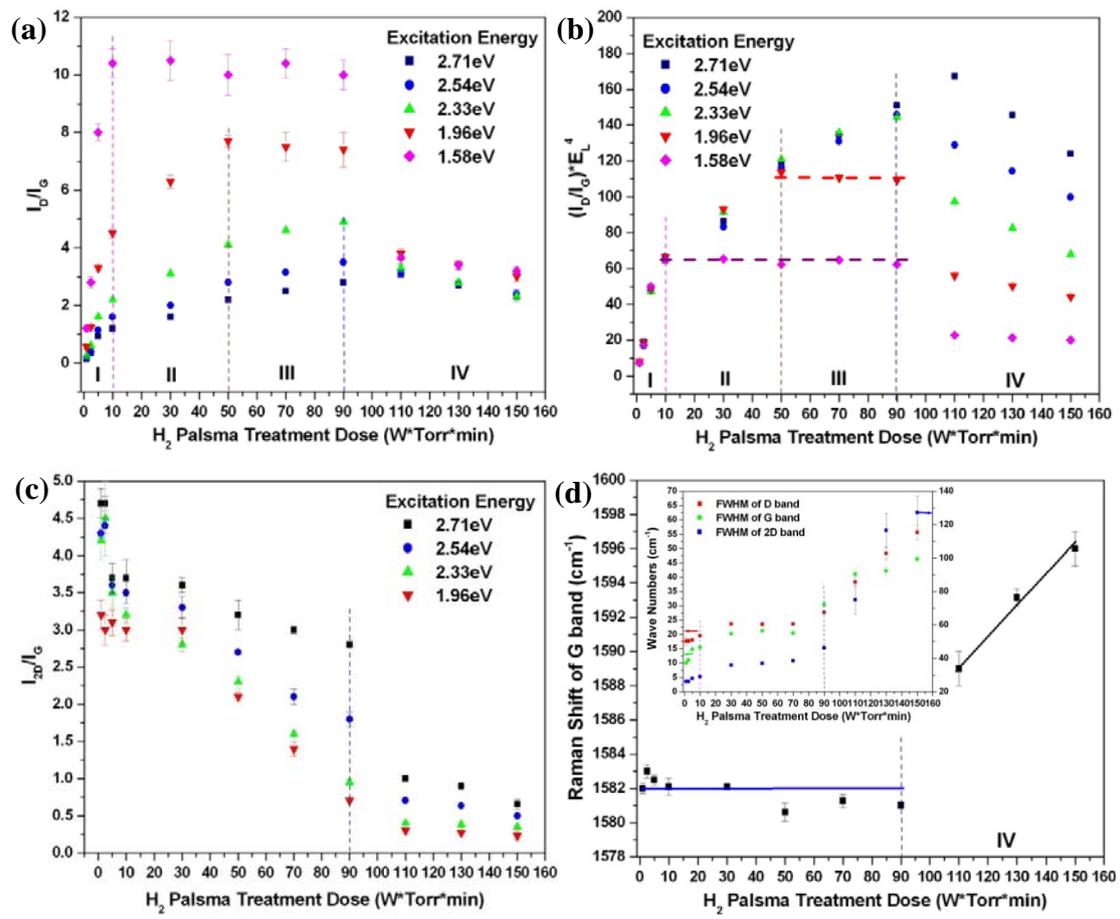

**Figure 2.**

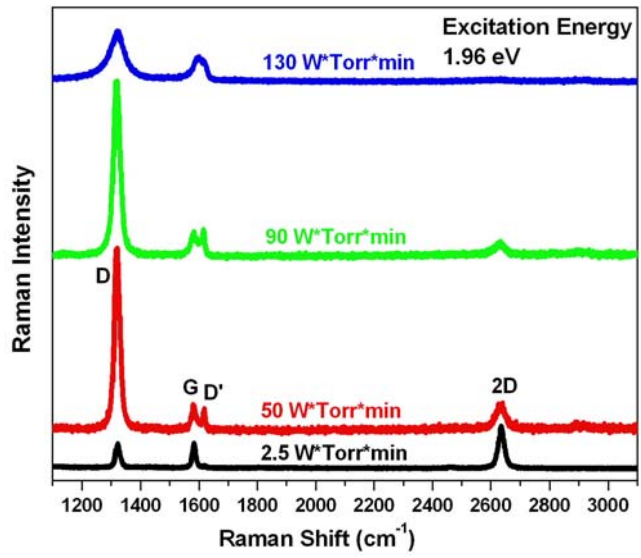

**Figure 3.**

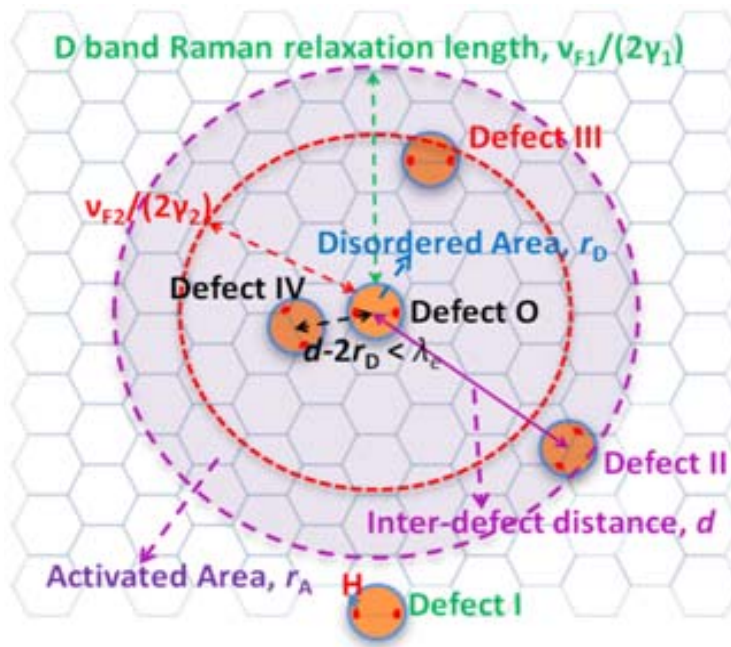

**Figure 4.**

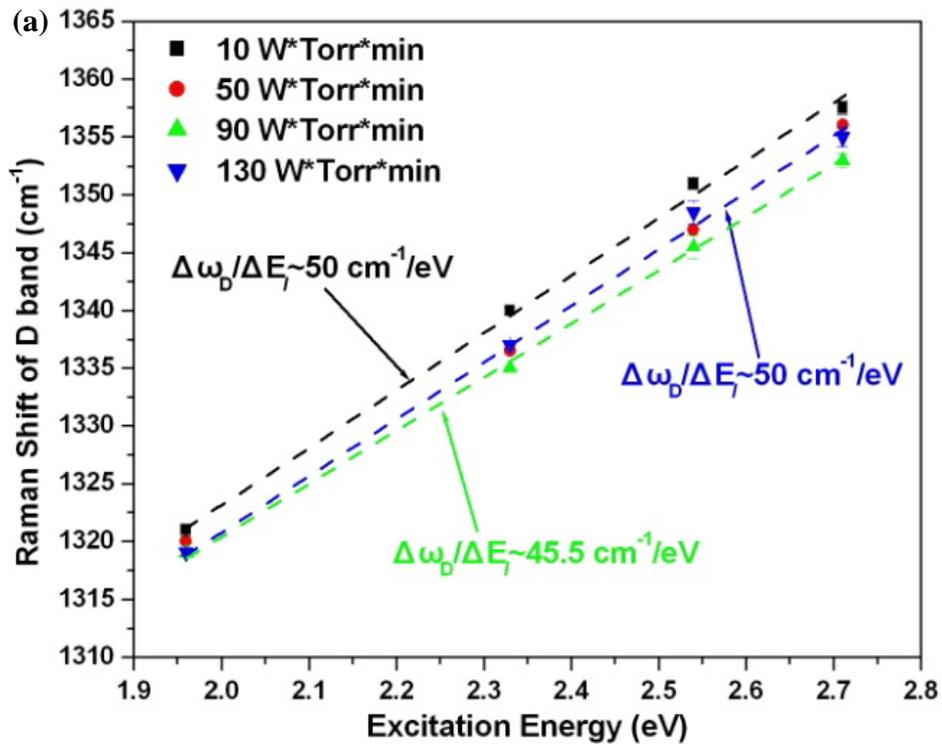

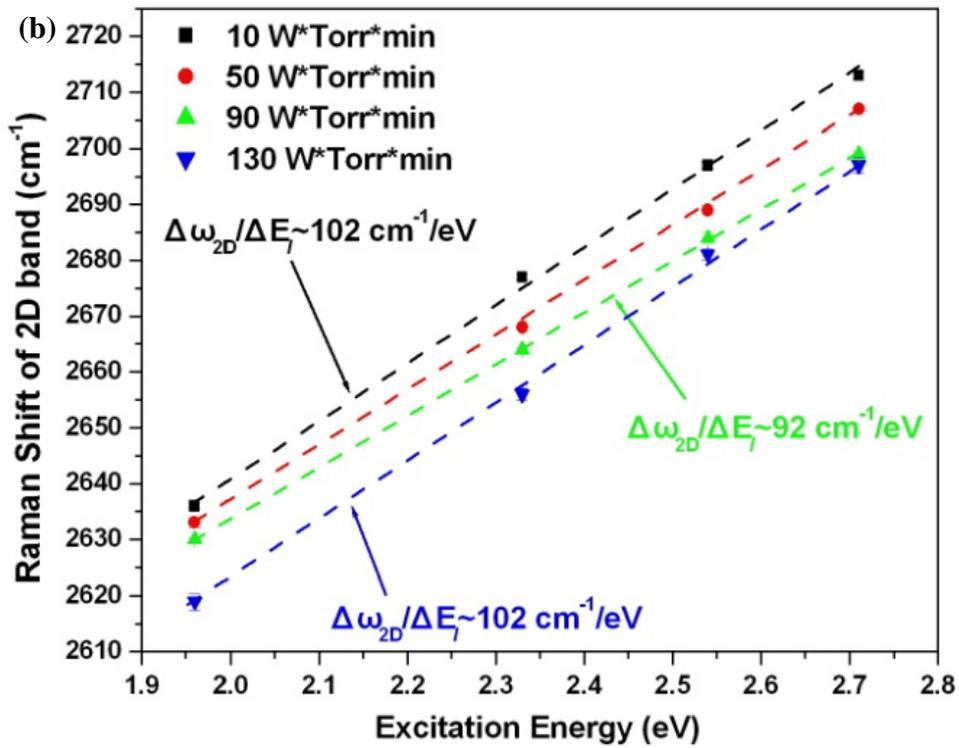

**Figure 5.**

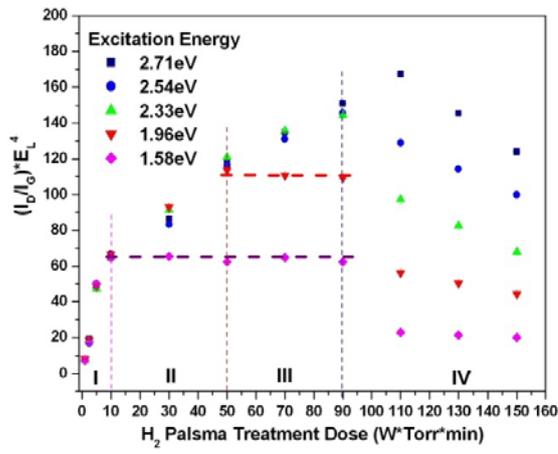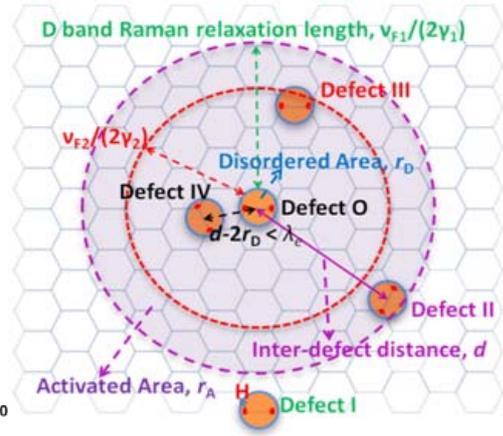

**TOC image**